\documentclass[prl,twocolumn,showpacs,amsmath]{revtex4}

\usepackage{graphicx}

\begin{document}

\title{
   Skyrmions in the Moore--Read state at $\nu=5/2$} 

\author{
   Arkadiusz W\'ojs$^{1,2}$,
   Gunnar M\"{o}ller$^1$,
   Steven H. Simon$^3$, and
   Nigel R. Cooper$^1$}

\affiliation{
       $^1$TCM Group, Cavendish Laboratory, 
           University of Cambridge,
           Cambridge CB3 0HE, United Kingdom\\
       $^2$Institute of Physics, 
           Wroclaw University of Technology,
           50-370 Wroclaw, Poland\\
       $^3$Rudolf Peierls Centre for Theoretical Physics, 
           University of Oxford,
           Oxford OX1 3NP, United Kingdom}

\begin{abstract}
We study charged excitations of the non-abelian Moore--Read liquid at 
filling factor $\nu=5/2$, allowing for spin depolarization. 
Using a combination of numerical studies, and taking account of non-zero 
well widths, we find that at sufficiently low Zeeman energy it is 
energetically favourable for charge $e/4$ quasiholes to bind into
``skyrmions'' of charge $e/2$. 
We show that skyrmion formation is further promoted by disorder, and argue 
that this can lead to a depolarized $\nu=5/2$ ground state in realistic 
experimental situations. 
We comment on the consequences for the activated transport.
\end{abstract}


\maketitle

The $\nu=5/2$ quantum Hall state \cite{Willett87,Eisenstein88} has 
been intensively studied in recent years, mostly due to accumulating 
evidence that it realizes a non-abelian phase of matter -- the 
Moore--Read state \cite{Moore91} -- which could serve as a platform 
for topological quantum computation \cite{Nayak08}. 
A crucial step toward identifying the experimental $\nu=5/2$ state as 
the Moore--Read phase \cite{Moore91,Read00,Moller08} is to demonstrate 
full spin polarization. 
Numerical simulations of model systems point strongly to a ground state 
that is fully polarized \cite{Morf98,Rezayi00,Feiguin09}. 
However, the experimental situation remains unclear \cite{Eisenstein88}. 
This issue has been reopened by recent experiments showing evidence 
of a depolarized ground state \cite{Pinczuk09} and emphasizing 
\cite{Dean08} the unusual parallel field dependence of the activation 
energy \cite{Eisenstein88}. 
Resolving this discrepancy has suddenly emerged as an important challenge.

In this Letter, we study spin polarization and charged excitations of 
the half-filled second Landau level (LL$_1$).
An important factor known to influence spin dynamics in LL$_1$ 
\cite{Cooper97} but remaining unexplored in the context of $\nu=5/2$ 
is the non-zero width $w$ of the quasi-two-dimensional electron layer. 
Indeed, the leading Coulomb pseudopotentials for a spin-singlet pair 
of electrons in LL$_1$ soften considerably (relative to spin-triplet) 
upon increasing $w$ to the order of a magnetic length 
$\lambda\equiv\sqrt{\hbar c/eB}$. 
The results of our exact diagonalization (ED) studies \cite{footnote-w} 
show that while the $\nu=5/2$ state remains spin-polarized, its charged 
excitations can become depolarized ``skyrmions'' at the experimentally 
relevant widths $w\gtrsim\lambda$. 

Skyrmions at $\nu=5/2$ are qualitatively different from those studied 
previously at $\nu=1$ or $1/3$ \cite{Sondhi93,MacDonald98}. 
Their charge is twice that of the fundamental excitation, $q=e/4$. 
Thus, the $2q$-charged skyrmion at $\nu=5/2$ can be viewed as a bound 
pair of like quasiparticles (QPs), held together against their Coulomb 
repulsion by a spin-texture. 
This binding of QPs into skyrmions has the striking consequence that 
skyrmion formation is {\em promoted} by disorder: disorder can act to 
trap and confine QPs to a potential well, thereby facilitating their 
binding into skyrmions. 
We show that this process can lead to a depolarized ground state under 
realistic experimental conditions. 
We suggest that this may account for the experimental observations of 
a reduction of polarization \cite{Pinczuk09}, and may play a role in 
the parallel field dependence of the activation energy 
\cite{Eisenstein88,Dean08}. 

\begin{figure}
\includegraphics[width=3.4in]{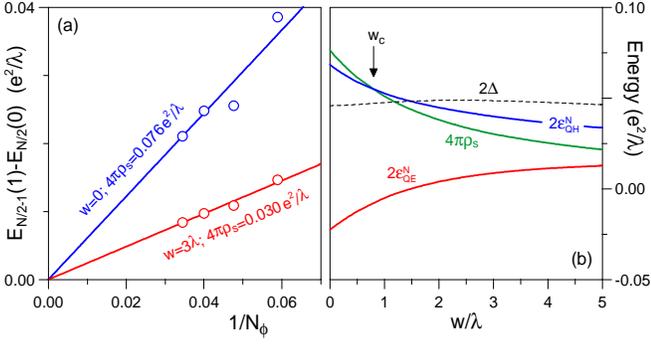}
\caption{(color online)
(a) Extraction of the spin stiffness $\rho_s$ at $\nu=5/2$ from 
the long-wavelength part of the spectra $E_S(L)$ with a single 
spin flip ($E$, $S$, and $L$ are the energy, spin, and angular 
momentum of $N\le16$ electrons on a sphere, at flux $N_\phi=2N-3$).
(b) Comparison of the skyrmion energy $4\pi\rho_s$ and the neutral 
QP energies $\varepsilon^{\rm N}$, establishing the skyrmion as the 
lowest positively charged excitation for sufficient layer widths 
$w$ ($\lambda$ is the magnetic length; $\Delta=\varepsilon_{\rm QH}
+\varepsilon_{\rm QE}$).}
\label{fig1}
\end{figure}

Skyrmions in (ferromagnetic) quantum Hall liquids are topological 
excitations in which the local spin orientation wraps once over the 
surface of the spin sphere. 
The coupling of spin and orbital degrees of freedom causes a skyrmion 
to appear as a single effective magnetic flux quantum, leading to a 
net charge $\nu'e$ where $\nu'$ is the filling fraction of the 
partially occupied LL \cite{Sondhi93}. 
(To be definite, we use the term ``skyrmion'' to denote the excitation 
with a net reduction of particle number, and ``anti-skyrmion'' for the 
excitation with a net increase.)\ 
At $\nu=5/2$ the partially occupied LL$_1$ is half-filled, so the 
skyrmion has charge $e/2$. 
For vanishing Zeeman energy, the energy of a large skyrmion is 
$\varepsilon_{\rm sky}=4\pi\rho_s$, and the spin stiffness $\rho_s$ 
can be found from the dispersion of a single spin wave. 
We used ED on a sphere to calculate the energy spectra of $N$ electrons 
in LL$_1$, with one spin flip, at the magnetic flux $N_\phi=2N-3$ 
corresponding to the Moore--Read polarized ground state. 
Long wavelength spin waves have quadratic dispersion, 
$E_{N/2-1}(L)=E_{N/2}(0)+8\pi N^{-1} \rho_s\, L(L+1)$, where $E_S(L)$ 
is the $N$-electron energy as a function of total spin $S$ and angular 
momentum $L$. 
As (for the numerically accessible $N\le16$) only the $L=1$ spin wave 
state falls below the continuum of polarized excitations, we estimate 
$\rho_s$ by regression of $E_{N/2-1}(1)$ vs.\ $N_\phi^{-1}$, as shown 
in Fig.~\ref{fig1}(a).

It is important to note that the skyrmion, or anti-skyrmion, has twice the 
charge of the elementary quasihole (QH), or quasielectron (QE). 
Therefore, in order to gauge stability, the skyrmion (anti-skyrmion) 
energy $\varepsilon_{\rm sky}$ must be compared to the energy of a 
QH (QE) pair created in the same polarized ground state by adding 
(removing) 
one flux quantum, but without causing depolarization. 
The relevant quantity in this context is the ``neutral'' QP energy 
$\varepsilon^{\rm N}$ \cite{Morf86,MacDonald86}, different from the 
(more common) ``proper'' energy $\tilde\varepsilon$ \cite{Morf86} 
by an appropriate fraction (here, $\pm1/8$) of the ground state 
energy per particle. 
Using the ground state and proper QP energies at $\nu=5/2$ 
\cite{Morf02,Wojs09} we find the width dependencies of 
$\varepsilon_{\rm QH}^{\rm N}$ and $\varepsilon_{\rm QE}^{\rm N}$ 
(with similar relations to the gap $\Delta=\varepsilon_{\rm QH}+ 
\varepsilon_{\rm QE}$ as reported earlier for other states in LL$_1$ 
\cite{MacDonald86}). 
The comparison of $\varepsilon_{\rm sky}(w)$ and 
$2\varepsilon_{\rm QP}^{\rm N}(w)$ in Fig.~\ref{fig1}(b) yields the 
prediction that a critical width $w_c$ of the order of $\lambda$ is 
sufficient to induce transition from QHs to skyrmions as the 
lowest energy positive excitations at $\nu=5/2$. 
In contrast, for all widths $w$ we have studied, our results show 
that the anti-skyrmion has higher energy than a QE pair.

Since QHs of the Moore--Read state exhibit non-abelian statistics, 
the binding of two QHs into a skyrmion allows for two distinct fusion 
channels (1 and $\psi$) \cite{Baraban09}. 
The case that we have described above -- in which the skyrmion has 
energy $4\pi\rho_{\rm s}$ -- refers to the fusion channel ``1'' in 
which the spin-texture consists of the Moore-Read ground state with 
no unpaired fermions. 
Pairing in the $\psi$-channel corresponds to a skyrmion spin-texture 
in the presence of an additional neutral fermion excitation (in a finite 
system, this arises for an odd number of particles). 
As we have confirmed by ED for $N\le20$, for skyrmion configurations
(with two bound QHs) this unpaired neutral fermion has a positive 
energy.
Thus, the fusion of QHs into a skyrmion has lower energy in the ``1'' 
channel. 
In view of this, we shall focus mainly on this fusion channel, using 
finite size studies with even $N$. 

The above considerations apply in the case of vanishing Zeeman 
energy $E_{\rm Z}$, when the skyrmion has infinite size. 
In order to understand the properties at $E_{\rm Z}>0$, we have 
conducted extensive ED studies of depolarized states at $\nu=5/2$, 
using the spherical geometry \cite{Haldane83}, with Coulomb 
repulsion modified by the layer width $w$ \cite{footnote-w}. 
An (infinite) skyrmion is represented on a sphere by a maximal 
spin-texture which uniformly covers its surface. 
In order to compute the ground states with the proper quantum 
numbers $L=S=0$ we used a projected Lanczos procedure, carried 
out in the subspace defined by $(L_z,S_z)$ but restricted to the 
relevant eigensubspace $(L,S)$. 
This enabled efficient generation of the spin eigenstates with 
dimensions reaching $1.4\times 10^9$ for $N=12$ and $N_\phi=26$. 

Skyrmions $\mathcal{S}^+\Phi$ (anti-skyrmions $\mathcal{S}^-\Phi$) 
can be anticipated at the values of flux $N_\phi$ one quantum above 
(below) a polarized ``parent'' ground state $\Phi$. 
For $\nu=5/2$ this parent could belong to the universality class 
of either the Moore--Read (Pfaffian) state $\Phi_{\rm MR}$ or its 
particle-conjugate (``anti-Pfaffian'') $\Phi_{\rm MR}^*$, on a 
sphere characterized by the ``shifts'' $\sigma\equiv2N-N_\phi=3$ 
and $-1$, respectively. 
This gives four possible skyrmion candidates $\Psi_\sigma$ 
identified as: 
$\Psi_{ 4}=\mathcal{S}^-\Phi_{\rm MR}$, 
$\Psi_{ 2}=\mathcal{S}^+\Phi_{\rm MR}$, 
$\Psi_{ 0}=\mathcal{S}^-\Phi_{\rm MR}^*$, and 
$\Psi_{-2}=\mathcal{S}^+\Phi_{\rm MR}^*$. 
Indeed, in the energy spectra $E_S(L)$ obtained for $N\le12$ 
(and for different widths $w\le3\lambda$) we find that only at 
those four shifts $\sigma=4$, 2, 0, $-2$ are the lowest $S=0$ 
states consistently nondegenerate ($L=0$). 
Furthermore, at each width $w$, their ground state energies 
per particle all extrapolate to the same value as that of their 
polarized parents. 

\begin{figure}
\includegraphics[width=3.4in]{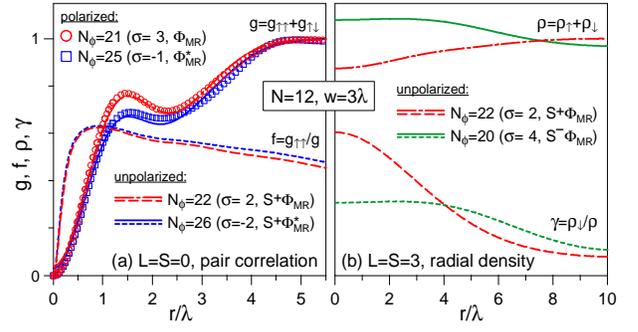}
\caption{(color online) 
(a) Charge and spin pair correlation functions ($g$ and $f$) 
of skyrmion states $\Psi_2=\mathcal{S}^+\Phi_{\rm MR}$ and 
$\Psi_{-2}=\mathcal{S}^+\Phi_{\rm MR}^*$ at $\nu=5/2$, compared 
to their Pfaffian parents, calculated for $N=12$ electrons and 
layer width $w=3\lambda$. 
(b) Single-electron charge and spin-flip densities ($\varrho$ 
and $\gamma$) of the finite-size skyrmions with an intermediate 
size $K=N/4$.}
\label{fig2}
\end{figure} 

Features of the correlation functions of $\Psi_\sigma$ fully 
support their interpretation as skyrmions. 
As shown on two examples in Fig.~\ref{fig2}(a), their charge 
pair correlation functions 
$g=g_{\uparrow\uparrow}+g_{\uparrow\downarrow}$ 
are virtually identical to those of their polarized parents. 
Moreover, an extended spin-texture is clearly revealed in a 
smoothly decreasing spin correlation function 
$f=g_{\uparrow\uparrow}/g$ (an initial increase of $f$ at short 
range being irrelevant due to the vanishing $g$). 
Our results for $\sigma=2$ are consistent with those of Morf 
\cite{Morf98}. 

Focusing now on $\Psi_2$, let us compare it to the following trial 
wave function of an ideal skyrmion. 
At $\nu=1$, the skyrmion wave function $\Psi_{\rm sky}$ can be written 
as the wave function $\Phi$ of a polarized filled LL times an additional 
factor describing the spin-texture \cite{MacDonald96}. 
The latter factor formally represents a many-body state of spinful 
bosons experiencing the (single quantum of) flux added to the $\nu=1$ 
parent, and it is uniquely defined by $L$ and $S$. 
Here, we extend this idea to a polarized parent at $\nu=5/2$, in the 
following chosen to be the Moore--Read state $\Phi_{\rm MR}$ 
\cite{footnote-qmc}. 
The skyrmion state $\Psi_{\rm sky}$ is constructed from $\Phi$ by 
attaching a unique spin-texture. 
The expression in the spherical coordinates $u_i$ and $v_i$ 
\cite{Haldane83} is fairly simple
\begin{equation}
   \Psi_{\rm sky}(\{u_i,v_i\})=
   \mathcal{P}\left[\Phi(\{u_i,v_i\})\times{u_i\choose v_i}\right].
\end{equation} 
Before projection $\mathcal{P}$ onto the global singlet $S=0$, this state 
describes a radial spin-texture and (like at $\nu=1$) combines eigenstates 
with different $L=S$. 
The squared overlaps of $\Psi_{\rm sky}$ with the unpolarized Coulomb ground 
state $\Psi_2$ were calculated in a standard way \cite{Moller09}. 
The values for $\Psi_{\rm sky}$ derived from $\Phi_{\rm MR}$ are given in 
Table~\ref{tab1}. 

We have also constructed an alternative trial skyrmion state 
$\Psi'_{\rm sky}$, as the ground state of a model Hamiltonian 
$H_{\rm sky}=V_0(0)+W_{3/2}(3)+\epsilon V_0(2)$, where $\epsilon\ll1$, 
and $V_S(m)$ and $W_S(m)$ are the pair and triplet pseudopotentials. 
The resulting state, $\Psi'_{\rm sky}$, is essentially the same as 
$\Psi_{\rm sky}$ defined above (squared overlap of 0.96 for $N=12$).
By tracing the successive action of each term we found that the (moderate) 
overlaps with $\Psi_2$ are largely lost at the stage of enforcing triplet 
correlations. 
This reflects the fact that the Pfaffian $\Phi_{\rm MR}$ is not a 
very accurate description of the Coulomb ground state $\Phi_3$ 
\cite{Rezayi00,Moller08}, here acting as the skyrmion's parent.
Indeed a comparison (Table~\ref{tab1}) of the overlaps 
of $\Psi'_{\rm sky}$ or $\Psi_{\rm sky}$ with the exact Coulomb 
skyrmion state shows that the reduction in total overlap is comparable 
to that between $\Phi_{\rm MR}$ and $\Phi_3$.

\begin{table}
\caption{
Squared overlaps of the trial skyrmion states $\Psi_{\rm sky}$ and 
$\Psi_{\rm sky}'$ and their Pfaffian parents $\Phi_{\rm MR}$ with 
the exact Coulomb ground states (unpolarized $\Psi_2$ and polarized 
$\Phi_3$) for $N=10$ electrons and layer widths $w=0$ and $3\lambda$.}
\begin{ruledtabular}
\begin{tabular}{cccc}
$w/\lambda$ & 
$|\left<\Psi_{\rm sky}|\Psi_2\right>|^2$ & 
$|\left<\Psi_{\rm sky}'|\Psi_2\right>|^2$ & 
$|\left<\Phi_{\rm MR}|\Phi_3\right>|^2$ 
\rule[-2mm]{0mm}{0mm}
\\
\hline 
0 & 0.51(3) & 0.5186 & 0.7016 \\
3 & 0.71(1) & 0.7394 & 0.8310
\end{tabular}
\end{ruledtabular}
\label{tab1}
\end{table}

We have also examined the partially polarized spectra at $\sigma=4$, 2, 
0, $-2$. 
Finite skyrmions of size $K\le N/2$ on the sphere have $L=S=N/2-K$ 
\cite{MacDonald96}. 
As shown on a few examples in Fig.~\ref{fig2}(b), the plots of 
single-particle charge and spin-flip densities 
($\varrho=\varrho_\uparrow+\varrho_\downarrow$ and 
$\gamma=\varrho_\downarrow/\varrho$) calculated in the lowest Coulomb 
states at $L_z=L=S_z=S=1$, 2, \dots\ indeed reveal accumulation of 
charge $2q$ and spin $K$ over a finite area $\propto K$ around a pole. 

We now turn to study the competition between the finite skyrmions and 
the spinless QPs. 
The analogy with $\nu=1$ or $1/3$ fails, as skyrmions at $\nu=5/2$ carry 
two charge quanta ($2q=e/2$), which allows their spontaneous break-up 
into pairs of repelling QPs. 
In order to resolve the issue of stability, we have carefully compared 
the skyrmion and 2QP energies, including the electrostatic corrections 
\cite{Morf02} to compensate for finite-size effects due to different 
spatial extent of the involved carriers. 
(At $S=0$, there is no correction as the charge distribution is uniform; 
at $S=N/2$ we assume, as standard, a concentrated charge;
a cubic interpolation is applied.)

\begin{figure}
\includegraphics[width=3.4in]{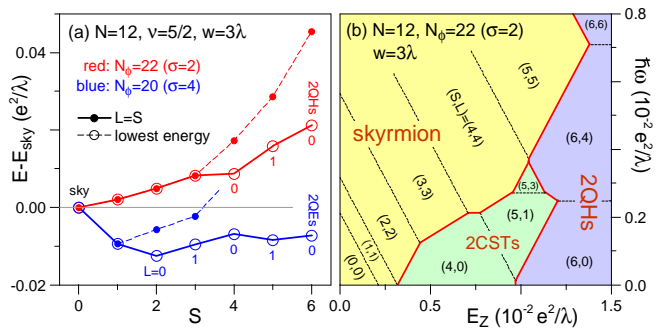}
\caption{(color online)
(a) Dependence of energy $E$ (counted from $E_{\rm sky}$ at $S=0$) 
on spin $S$, for $N=12$ electrons at flux $N_\phi=20$ and 22. 
Solid and dashed lines connect the ground states and the lowest 
$L=S$ states at each $S$. 
The energies include electrostatic corrections defined in the text. 
(b) Phase diagram in the plane of Zeeman energy $E_{\rm Z}$ and 
lateral harmonic confinement $\hbar\omega$, showing transitions 
between various states: two $e/4$-charged quasiholes (QHs); 
two charged spin-textures (CSTs); or one skyrmion ($L=S$).}
\label{fig3}
\end{figure}

As illustrated in Fig.~\ref{fig3}(a) for $N=12$ and 
$w=3\lambda$, the skyrmion/anti-skyrmion asymmetry is very strong 
at $\nu=5/2$. 
In agreement with Fig.\ref{fig1}, large skyrmions are energetically 
favoured over the QHs, but large anti-skyrmions have higher energy 
than the QEs. 
As expected, the finite skyrmions characterized by $L=S$ are no longer 
the lowest states when spin polarization becomes too high. 
The emergence of lower-energy states at $K\lesssim N/4$ signals the 
break-up of a skyrmion into two separate objects (whose counter-aligned 
angular momenta give rise to the observed oscillation between $L=0$ and 1). 
At full polarization, the two objects are Moore--Read QHs; below that, 
they are $q$-charged spin-textures (CSTs) formed around the individual 
QHs \cite{Dimov08,footnote-odd}.

For a clean (disorder-free) sample, we find that, while the $\nu=5/2$ 
ground state remains polarized at $E_{\rm Z}=0$, the nature of its charged 
excitations (in sufficiently wide wells) depends on $E_{\rm Z}$. 
For large $E_{\rm Z}$ the activation gap is set by the energy of a single 
(polarized) QE-QH pair. 
For small $E_{\rm Z}$, these are replaced by CSTs. 
The minimal activation gap is never expected to involve a skyrmion, 
as this requires creation also of two $-e/4$ QEs (or CSTs). 

Skyrmions can become relevant in a clean sample when dilute QHs are 
introduced into the ground state by tuning to $\nu<5/2$. 
From Fig.\ref{fig3}(b), dilute QHs will convert to CSTs for 
$E_{\rm Z}\lesssim 0.01\,e^2/\lambda$ and then pair up into finite-sized 
skyrmions for $E_{\rm Z}\lesssim0.003\,e^2/\lambda$. 
At $\nu>5/2$, dilute QEs may convert into CSTs at small $E_{\rm Z}$; 
however they are not expected to bind into anti-skyrmions. 
Since the skyrmion is formed by binding two QHs, its stability may be 
\emph{enhanced} by a non-zero QH concentration. 
Thus, with increasing the QH density at fixed $E_{\rm Z}$, a Wigner 
crystal of QHs could convert into a Wigner crystal of skyrmions. 

Real experimental samples always involve some disorder. 
This can lead to skyrmion formation in the ground state. 
Even at the centre of the $\nu=5/2$ plateau, long-range disorder can 
nucleate puddles of QHs and QEs, which can evolve into CSTs or skyrmions 
at small $E_{\rm Z}$. 
In states like $\nu=1$ or $1/3$, disorder acts to disfavour skyrmions, 
as it reduces the size of trapped carriers. 
Surprisingly, the effect of disorder at $\nu=5/2$ is the opposite: 
it can act to bring two repelling QHs together and therefore promote 
their merging into a skyrmion. 
In order to investigate these effects, we have studied two QHs in 
a lateral harmonic confinement of strength $\hbar\omega$, which models 
the potential in a local minimum of the disorder. 
From the ED results for the ground state energy $E_S(L)$, we construct 
the phase diagram shown in Fig.~\ref{fig3}(b). 
This shows that, as disorder strength ($\hbar\omega$) is increased, 
the skyrmion is stable up to larger values of $E_{\rm Z}$. 

Our results show that, for GaAs samples where $\nu=5/2$ occurs at
$B\lesssim 6 \mbox{T}$, QHs may bind into skyrmions if assisted by
disorder.  
This follows from Fig.\ref{fig3}(b), where the reduction of spin from 
its maximal value occurs for $E_{\rm Z}\lesssim0.014\,e^2/\lambda$.  
However, we emphasise that this numerical value is subject to significant 
uncertainty owing to finite size effects.
Still, our results make the clear qualitative point that, even at the
centre of the $\nu=5/2$ plateau, in samples with sufficiently low
Zeeman energy, the ground state will not be fully polarized: QHs
trapped by disorder will bind into skyrmions causing depolarization.
We suggest this could account for the experimental evidence of the
lack of spin polarization at $\nu=5/2$ \cite{Pinczuk09}.

The appearance of this unpolarized ground state may have consequences 
for the activated transport. 
The situation is very different from $\nu=1$ or $1/3$, where skyrmions 
are the charged carriers and increasing $E_{\rm Z}$ increases the 
activation gap. 
At $\nu=5/2$, skyrmions can form in a (disordered) {\em ground state}.
Activated transport will still involve the motion of $e/4$ excitations 
(the minimal charged object), but now hopping of QHs will have an 
activation energy that includes a contribution to overcome their binding 
into skyrmions, which {\em decreases} with increasing $E_{\rm Z}$. 
This effect could potentially account for the observed reduction of 
activation energy with {\em small} in-plane fields \cite{Eisenstein88,Dean08}.
We further note that the tendency toward binding of topologically nontrivial 
QHs into topologically trivial skyrmions could strongly affect interference 
experiments designed to probe non-abelian statistics \cite{Willett09}.

In conclusion, we have shown that for realistic widths $w\gtrsim\lambda$ 
the QH excitations of the $\nu=5/2$ Moore-Read state can bind to form 
skyrmions at small Zeeman energy. 
Long-range disorder acts to bring trapped QHs together and to promote 
skyrmion formation. 
This can lead to a depolarized ground state even at the centre of 
the plateau. 
This may account for recent measurements of spin-depolarization 
\cite{Pinczuk09}, and for the unusual in-plane field dependence of 
the activation energy \cite{Eisenstein88,Dean08}. 

We acknowledge the support of the EU under the Marie Curie grant 
PIEF-GA-2008-221701 and the Polish MNiSW under grant N202-071-32/1513 
(A.W.); Trinity Hall (G.M.); and EPSRC under EP/F032773/1 (N.R.C.).


\begin{thebibliography}{99}
\bibitem{footnote-w}
We assume hard wall confinement in the perpendicular direction, 
yielding single-particle density $\propto\cos^2(\pi z/w)$.
\bibitem{footnote-qmc}
The same construction could be generalized to other polarized states, 
e.g.\ an optimized weakly paired state \cite{Moller08}, or the numerical 
Coulomb ground state $\Phi_3$.
\bibitem{footnote-odd}
A comparison of the ED results for the unpolarized states with $\sigma=2$, 
for even and odd $N$ ($S=0$ and $1/2$, respectively), suggests that the 
QH binding at small $E_{\rm Z}$ occurs in the ``1'' fusion channel also 
for small skyrmions.
\end{thebibliography}
\end{document}